\documentstyle[12pt]{ioplppt}
\input epsf.tex

\def\paper#1#2#3#4#5{#1, {#3} {\bf #4}, \rm #5 (#2).}
\def\PRB{\it Phys. Rev. B}
\def\PRE{\it Phys. Rev. E}

\begin{document}
\jl{1}

\title{Quenched bond dilution in two-dimensional Potts models}

\author{Christophe Chatelain\dag, Bertrand Berche\ddag\ and 
Lev N. Shchur\S\ddag}
\address{\dag\ Institut f\"ur Theoretische Physik, Universit\"at Leipzig,\\ 
D-04109 Leipzig, Germany}
\address{\ddag\ Laboratoire de Physique des Mat\'eriaux\ftnote{6}{Unit\'e de
Mixte de Recherche  CNRS No 7556}, Universit\'e Henri
Poincar\'e, Nancy 1\\
B.P. 239, F-54506 Vand\oe uvre les Nancy, France}
\address{\S\ Landau Institute for Theoretical Physics, \\ 
Russian Academy of Sciences, \\ 
Chernogolovka 142432, Russia}

\date{\today}

\begin{abstract}
\baselineskip=16pt
We report a numerical study of the bond-diluted 2-dimensional Potts model 
using transfer matrix calculations. 
For different numbers of states per spin, we show that the critical
exponents at the random fixed point are the same as in  
self-dual random-bond cases. In addition, we determine the multifractal 
spectrum associated with the scaling dimensions of the moments of the 
spin-spin correlation function in the cylinder geometry. We show that the 
behaviour is fully compatible with the one observed in the random bond case,
confirming the general picture according to which a unique fixed point 
decsribes the critical properties  of different classes of disorder:
dilution, self-dual binary random-bond, self-dual continuous random bond.
\end{abstract}

\maketitle

\section{Survey of theoretical and experimental results in $2D$}
\label{sec:intro}
Quenched disorder coupled to the energy-density has
been the subject of an intensive activity in statistical physics, essentially
in two-dimensional ($2D$) systems, during the past decades. 
The qualitative influence of 
disorder at {\em second order} phase 
transitions is well understood since Harris proposed a celebrated
relevance criterion~\cite{Harris74}. 
At {\em first order} transitions, randomness obviously
softens the transitions, and, under some circumstances may even induce second 
order transitions according to a picture first proposed by Imry and 
Wortis~\cite{ImryWortis79} and then stated on more rigorous grounds 
by Aizenman 
and Wehr~\cite{AizenmanWehr89,HuiBerker89}, implying an important result 
that an infinitesimal disorder induces  continuous transitions in $2D$.
The $q-$state Potts model~\cite{Wu82} is the 
natural candidate for the investigations of influence of disorder in a 
perspective linked to critical phenomena, since the pure model exhibits
two different regimes: a second order phase transition when $q\leq 4$  and
a first order one for $q>4$ (in $2D$). Many results were obtained in 
both regimes for self-dual quenched randomness in this model in the last ten 
years,
including perturbative 
expansions~\cite{Ludwig87,LudwigCardy87,Ludwig90,DotsenkoPiccoPujol95a,DotsenkoDotsenkoPiccoPujol95,JugShalaev96,DotsenkoDotsenkoPicco97,Lewis98}, 
Monte Carlo 
simulations~\cite{Picco98,ChatelainBerche98a,ChatelainBerche98b,PalagyiChatelainBercheIgloi99,OlsonYoung99} 
or Transfer Matrix calculations~\cite{Glaus87,CardyJacobsen97,Picco97,JacobsenCardy98,ChatelainBerche99,ChatelainBerche00}, high-temperature series 
expansions~\cite{RoderAdlerJanke98,RoderAdlerJanke99} 
or recently short-time
dynamic scaling~\cite{PanEtal01}. The numerical 
studies showed that many difficulties, like
the lack of self 
averaging~\cite{Derrida84,AharonyHarris96,WisemanDomany98a,WisemanDomany98b} 
or varying effective exponents due to crossover phenomena
may occur. Averaging physical quantities over the samples 
with a poor statistics 
may thus lead to erroneous determinations of the critical exponents. 
We also note that the previoulsy mentioned studies were reported
in the case of the random bond system with self-dual probability distributions
of the coupling strengths in order to preserve the exact knowledge of the
transition line.

In real experiments on the other hand, disorder is inherent to the
working-out process and may result e.g. from the presence of impurities or
vacancies in a sample produced in Molecular Beam Epitaxy or
sputtering experiments. For the description of such a disordered system, 
dilution is thus more realistic than for example a random distribution of
non-vanishing couplings. 
Since universality is expected to hold
in non frustrated random systems, the detailed structure of the Hamiltonian
should not play any determining role in universal quantities like critical 
exponents,
but dilution  presumably produces a quite strong disorder compared to e.g.
a binary random-bond distribution of coupling strengths, and crossover 
phenomena may alter the determination of the universality class.

Experimentally, the role of disorder in $2D$ systems has been investigated
in several systems.
Illustrating the influence of random defects in the case of the 
$2D$ Ising model universality class, samples made
of thin magnetic amorphous layers of 
(Tb$_{0.27}$Dy$_{0.73}$)$_{0.32}$Fe$_{0.68}$ of 10~\AA\  width, 
separated by non magnetic spacers of 100~\AA\  Nb in
order to decouple the magnetic layers were 
produced using sputtering techniques. A structural analysis (high resolution
transmission electron microscopy and $x$-ray analysis) was performed to
characterize the defects inherent to such amorphous structures (types of voids
or possibly compositional disorder), and in spite of these random defects 
separated on average by a distance of a few nm, the samples were 
shown to exhibit Ising-like 
singularities with critical exponents 
$\beta=0.126(20)$, $\gamma=1.75(3)$ and $\delta=15.1(10)$~\cite{MohanEtal98}.
This is coherent with the fact that disorder does not change the universality
class of the $2D$ Ising model, apart from logarithmic corrections
which are not easy to observe experimentally, since their role 
becomes prominant only in the very neighbourhood of the critical point.
For this reason, the Ising model is probably not the best system to study 
quantitatively the influence of randomness experimentally.

More interesting from the point of view of critical phenomena is the case
of a  beautiful experimental confirmation 
of the Harris criterion reported in a Low Energy Electron Diffraction  
(LEED) investigation of a 
$2D$ order disorder 
transition~\cite{SchwengerEtAl94} belonging to the
4-state Potts model universality class. Order-disorder transitions of 
adsorbed atomic layers are known to belong to different two-dimensional 
universality classes depending on the type of superstructures in the ordered 
phase of the adlayer~\cite{DomanyEtAl78,Rottman81}. 
The substrate plays a major role in adatom ordering,
as well as the coverage (defined as the number of adatoms per surface atom)
which determines the possible superstructures of the overlayer.
For example, sulfur chemisorbed on 
Ru$(001)$ exhibits four-state or three-state Potts critical singularities
for the $p(2\times 2)$ and the $(\sqrt 3\times\sqrt 3)R30^\circ$
respectively~\cite{SokolowskiPfnur94} (at coverages $1/4$ and $1/2$). 
Other examples can be found in the 
literature, for example 
oxygen on Ru$(0001)$
ordered in a $p(2\times 2)$~\cite{PiercyPfnur87} or $p(2\times 2)$-Cs or K 
on Cu$(111)$~\cite{FanEtAl89} 
also belong to the four-state Potts
universality class, while $(\sqrt 3\times\sqrt 3)R30^\circ-1$CO has
three-state Potts singularities~\cite{OverEtAl97}.
The case of the $(2\times 2)-$2H/Ni$(111)$ order-disorder transition of
hydrogen adsorbed on the $(111)$ surface of Ni thus belongs to the
$2D$ four-state Potts model universality class (with expected exponents
$\beta=1/12\simeq 0.083$, $\gamma=7/6\simeq 1.167$, and $\nu=2/3\simeq 0.667$
for example). 

Using LEED, it is 
possible to measure these exponents through the diffracted intensity
$I({\bf q})$ or structure factor. This is the two-dimensional Fourier transform
of the pair correlation function of adatom density, where long range 
fluctuations produce an isotropic Lorentzian 
centered at the superstructure 
spot position  ${\bf q}_0$ with a peak intensity given by the
susceptibility and a width determined by the inverse correlation length, while
long range order gives a background  proportional to the order 
parameter squared:
\begin{equation}
	I({\bf q})=\langle m^2\rangle\delta({\bf q}-{\bf q}_0)
	+\frac{\chi}{1+\xi^2({\bf q}-{\bf q}_0)^2}.
\end{equation}
The following exponents were thus 
measured~\cite{SchwengerEtAl94,BuddeEtAl95,VogesPfnur98}
$\beta=0.11\pm 0.01$, $\gamma=1.2\pm 0.1$ and $\nu=0.68\pm 0.05$ in correct 
agreement with 4-state Potts values (the small deviation, especially for the
exponent $\beta$, is possibly attributed to the logarithmic corrections
to scaling of the pure 4-state Potts model~\cite{CardyNauenbergScalapino80}).
The same experiments were then reproduced in the presence
of  intentionally added oxygen
impurities, at 
a temperature which is above the ordering temperature of pure oxygen
adsorbed on the same substrate. The mobility of these oxygen atoms is
furthermore considered to be low enough at the hydrogen order-disorder
transition critical temperature that they essentially represent quenched
impurities randomly distributed in the hydrogen layer, and the new measured
exponents become $\beta=0.135\pm 0.010$, $\gamma=1.68\pm 0.15$ and 
$\nu=1.03\pm 0.08$. It definitely rules out the role of these
oxygen impurities as extended defects, like steps for example, which,
according to Ref.~\cite{SchwengerEtAl94},
produce a rounding of the transition by a simple 
finite-size-scaling effect. The modification of the universality class
in the presence of quenched disorder is understood with Harris
criterion which predicts such a situation when the exponent $\alpha$ of the 
specific heat is positive for the pure system
($\alpha=2/3$ for the 4-state Potts model).

The aim of this paper is to perform 
numerical simulations of the bond diluted Potts model for several values 
of the number of states per spin (in order to cover the two
different regimes of the pure system's phase transitions) and provide a 
reliable comparison of the diluted and
random-bond problems. This will be achieved through a systematic
comparison with previous results obtained for self-dual random-bond
disorder.
Here we stress that self-duality is an internal symmetry which can lead to
some simplifications that the dilute problem does not present and thus
it is worth comparing both problems.
In section~\ref{sec:algo}, the details of the numerical techniques
are summarized. This section also provides an exposition of the different 
extrapolation techniques in order to leave the discussion of the physical
results to the following of the paper. 
Section~\ref{sec:PhaseDiag} presents the phase diagram of
the diluted Potts models with $q=3$, 4, and 8, and section~\ref{sec:TM}
the results of transfer
matrix calculations and the critical behaviour of the
order parameter correlation function.

\section{Methodology and numerical techniques}
\label{sec:algo}
\subsection{Definition of the model}
In this paper, we study the 2-dimensional diluted $q$-state Potts model 
defined by the 
following Hamiltonian :
	\begin{equation}
	-\beta{\cal H}=\sum_{(i,j)} K_{ij}\delta_{\sigma_i,\sigma_j}
	\end{equation}
where the sum is restricted to nearest neighbours on a square lattice, the
degrees of freedom $\{\sigma_i\}$ can take $q$ values and the exchange
couplings $\{K_{ij}\}$ are quenched  independent random variables 
chosen according to a
binary distribution of non vanishing and vanishing values
	\begin{equation}
	{P}(K_{ij})=p\ \!\delta(K_{ij}-K)+(1-p)\delta(K_{ij}).
	\end{equation}
The value $p=1$ corresponds to the pure system and $p_c=1/2$ 
is the bond percolation threshold. Below this threshold there cannot exist any
ordered state at finite temperature and the phase transition temperature
vanishes. 
\subsection{Transfer matrix study}
The system is first studied using the 
transfer matrix method introduced by Bl\"ote and 
Nightingale~\cite{blotenightingale82}, which takes
advantage of the Fortuin-Kasteleyn representation~\cite{fortuinkasteleyn69}
in terms of graphs of the
partition function of the Potts model in order 
to reduce the dimension of the Hilbert 
space. In the Fortuin-Kasteleyn representation, the transfer 
matrix (with no magnetic
field)
\begin{equation}
	Z={\rm Tr}\prod_{(i,j)}
	(1+\delta_{\sigma_i,\sigma_j} u_{ij}),
\end{equation} 
where $u_{ij}={\rm e}^{K_{ij}}-1$, is expanded as a
sum over all the possible graphs $g$ (with $s$ sites, $l(g)$ loops and
$c(g)$ independent clusters) leading to the random cluster model:
\begin{equation}
	Z=q^s\sum_g q^{l(g)}\prod_{(i,j)/b_{ij}=1}
	\left(\frac{u_{ij}}{q}\right),
\end{equation} 
$b_{ij}$ being the bond variables.
Bl\"ote and Nightingale suggested to introduce a set of connectivity states
which contain the information about which sites on a given row
belong to the same cluster when they are interconnected through a part of
the lattice previously built. A unique connectivity label
$\eta_{i}=\eta$ is attributed to all the sites $i$ of such a cluster. 
In the connectivity space, $|Z(m)\rangle$ is a vector
whose components are given by the partial partition function 
$Z(m,\{\eta_i\}_m)$
of a strip of length $m$ whose connectivity on the last row is given by
$\{\eta_i\}_m$. The connectivity transfer matrix is then defined
according to $|Z(m+1)\rangle={\bf T}_m|Z(m)\rangle$ and the partition 
function of a strip of
length $m$ becomes $|Z(m)\rangle=\prod_{k=1}^{m-1} {\bf T}_k|Z(1)\rangle$,
where $|Z(1)\rangle$ is the statistics of uncorrelated spins.

\subsubsection{Free Energy Density}

The quenched free energy density is given by the  Lyapunov exponent of the
product of an infinite number of transfer matrices
${\bf T}_k$~\cite{furstenberg63}
\begin{equation}
	\overline{f_L}=-L^{-1}\Lambda_0(L).
\label{eq-F-ave}
\end{equation}	
\begin{equation}
	\Lambda_0(L)=\lim_{m\to\infty}\frac{1}{m}
	\ln\left|\!\left|\left(\prod_{k=1}^m 
	{\bf T}_k\right)
	\mid\! v_0\rangle\right|\!\right|,
\label{eq-Furst}
\end{equation}	
where $\mid\! v_0\rangle$ is a unit initial vector. 
In order to determine the error
induced by the truncation of this product to a finite number of terms, we
studied the fluctuations of the free energy density with respect to the number 
of iterations of the transfer matrix. The tests have only been performed for
the 4-state Potts model on strips of width $L=8$ with a pseudo critical  
exchange coupling $K_c=1.16215$ at the dilution $p=0.75$. 
This value of the exchange coupling is a 
rough estimate of the critical exchange coupling at this dilution as it
will be shown later.

\begin{figure}
	\epsfysize=10.5cm
	\begin{center}
	\mbox{\epsfbox{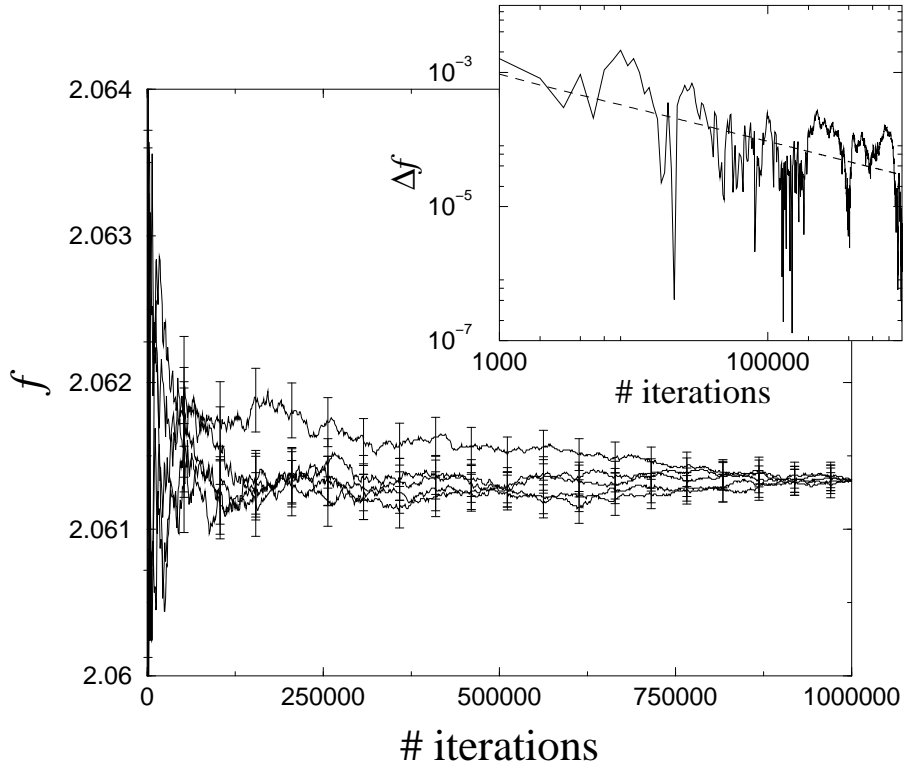}}
	\end{center}\vskip 0cm
	\caption{Convergence of the free energy density (for five different
		disorder realisations) as a function of the number of 
		iterations of the transfer matrix.
 		For the sake of clarity, error bars 
		are only displayed every 50000 iterations. 
		The parameters take the following values
		$L=8$, $p=0.75$, $K_c=1.16215$. The inset shows
		statistical fluctuations 
		$\Delta f=\sqrt{\overline{f^2}-\bar f^2}$  of 
		free energy density with respect
		to the number of iterations. The dashed
		line (which is a guide for the eyes) is a power-law with 
		exponent $-1/2$.}
	\label{FigConvergence-f}
\end{figure}
\begin{figure}[ht]
	\epsfysize=10cm
	\begin{center}
	\mbox{\epsfbox{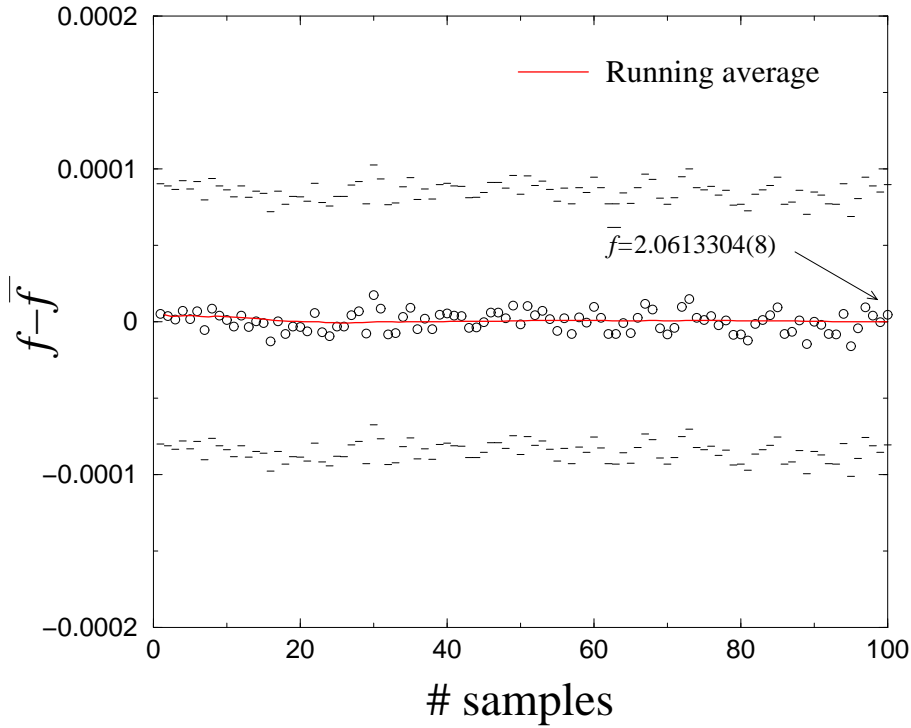}}
	\end{center}\vskip 0cm
	\caption{Convergence of the average of the free energy density $f$
		obtained by iterating $10^6$ times the transfer matrix with
		respect to the number of disorder realisations entering the
		average. Each circle corresponds to the free energy density
		of a given sample and the dots to the upper and
		lower bounds of its error bar. The solid line is the running
		average and the final estimate $\langle f\rangle$ is written 
		at its right edge. The parameters take the following values
		$L=8$, $p=0.75$, $K_c=1.16215$.}
	\label{FigConvergence-fAvg}  
\end{figure}

In figure~\ref{FigConvergence-f}, we present the estimates of the free energy
density for five independent runs and the inset shows its standard deviation.
For a self-averaging quantity $X$, the reduced standard deviation 
squared should be proportional to the inverse volume of the system,
$R_X=(\overline{X^2}-\bar X^2)/\bar X^2\sim 1/LN$ (the number of iterations
$N$ of the transfer matrix is the length of the strip), and thus at fixed
strip width $L$, $\Delta X\sim 1/\sqrt N$.
The standard deviation of the free energy density indeed exhibits an inverse 
square root decay  $1/\sqrt N$ 
with the number $N$ of iterations of the transfer
matrix (see inset in figure~\ref{FigConvergence-f}), 
i.e. it reveals that the estimates of the free energy density 
are weakly correlated from row to row and the free energy is self-averaging. 
It thus turns out to be preferable to average the
estimates of the free energy density obtained by different runs 
rather than using very long strips for which one would
accumulate truncation errors.
Figure~\ref{FigConvergence-fAvg} shows such an example of average of the free
energy density estimated by independent runs. Despite the fact that the
standard deviation of the free energy density seems to over-estimate the
true error by at least one order of magnitude, this definition of the error is
a safe choice and will be kept in the following. Up to now, all runs have been 
performed using $100$ independent runs of $10^6$ iterations of the transfer 
matrix, leading to an accuracy of 6-7 digits in the free energy density.

\subsubsection{Central charge}

For a pure system, the central charge $c$ is defined as the universal 
coefficient in the lowest-order correction to scaling of
the free energy density ${f_L}$:
	\begin{eqnarray}
	    {f_L}=f_\infty-L^{-2}\left[{\pi c\over 6}\right.&
		+b_\omega L^{-\omega}+b_{2\omega} L^{-2\omega}+\dots\nonumber\\
		&\left.+a_2L^{-2}+a_4L^{-4}+\dots\phantom{a\over b}\right],
	\end{eqnarray}	
where the regular contribution is 
\begin{equation}
	f_\infty=\lim_{L\rightarrow +\infty} {f_L}
\end{equation}
and $-\omega$ is the exponent associated to the irrelevant vacancy 
field~\cite{Nienhuis82,BCN86,Reinicke87,deQueiroz00}.
For a disordered system, $c$ is defined in the same way from the finite-size
behaviour of the quenched average free energy density $\overline{f_L}$, and
numerically, since the strip widths available are small, we can only expect
to measure effective central charges which depend on the dilution,
$c_{eff}(p)$, and which would converge towards the true value $c$ in the
thermodynamic limit.
One obviously expects the existence of higher order corrections to scaling, but
since no analytical expression is known, we cannot include explicite
size dependence for higher order terms and the possible corrections are 
taken into account by
fitting the free energy density including a $1/L^4$ non universal 
correction~\cite{DotsenkoJacobsenLewisPicco99,JacobsenPicco00}
	\begin{equation}
	    \overline{f_L}=f_0-{\pi c_{eff}\over 6L^2}
		+a_2L^{-4}.\label{eq-fbar} 
	\end{equation}
Fits with polynomials in $1/L^2$ of degrees ranging from 2 to 4 (i.e. 
up to $L^{-8}$) were
also considered, but do not increase the accuracy\footnote{This choice is 
arbitrary and has no theoretical grounds, but since we do not know the exact
expansion, a polynomial fit mimics the effective expansion for the
small strip widths available in the numerical computations.
After several trials, it turns out that the most
numerically stable estimates of $c_{eff}$ are those obtained 
with a polynomial of 
degree $2$ (including only the $L^{-4}$ term) in the range of lattice sizes 
$[3:8]$. The addition of a $L^{-6}$ term in the same range
$L\in[3:8]$ gives in the vicinity of the maximum the same estimate of the
central charge within numerical accuracy, i.e. the terms of order higher than
$1/L^4$ can be neglected. Polynomials of degree $4$ lead to non reliable
results, with error bars of
order $\Delta c_{eff}\sim 0.1$ due to the small number of degrees of freedom.}.

\subsubsection{Improvement of the calculation of the central charge}
\label{subsubsec:improvement}

\begin{figure}[ht]
        \epsfysize=10cm
        \begin{center}
        \mbox{\epsfbox{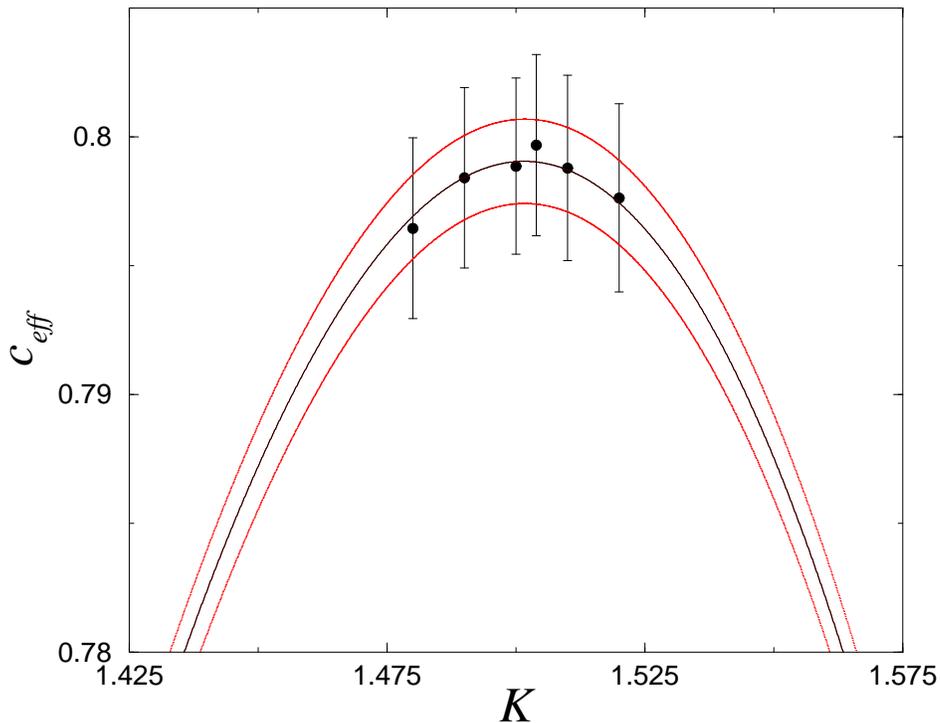}}
        \end{center}\vskip 0cm
        \caption{Central charge $c_{eff}$ of the 3-state Potts 
	model at dilution
        $p=0.75$ versus the exchange coupling $K$ as obtained with
        independent runs of 100 samples (symbols) and with a semi-analytic
        calculation with 500 samples of $10^6$ iterations of the transfer
        matrix (solid curve). The ditted lines correspond to the up and down
        limits of the errors bars of the semi-analytic calculation.}
        \label{FigChargeCentralePoly-q3p.75}
\end{figure}

In order to obtain more accurate estimates of the critical temperature,
especially at small  values of the number of states $q$ where the slow
variation of the effective central charge makes this 
determination difficult, we have
used a {\em  semi-analytical} transfer matrix. 
Instead of computing the product of the 
transfer matrices at a given value of the parameter $u=e^K-1$, the
calculations are performed at $u+\delta u$. All elements of the transfer
matrices are expressed as polynomials of $\delta u$. This parameter 
shift $\delta u$ is supposed small and the polynomials 
are truncated to the second order in $\delta u$. The free energy 
and its error bar are thus expressed as polynomials of $\delta u$ 
too and can be numerically computed over a continuous range of parameters
centered around $u$. The central charge is then extrapolated using the fitting 
procedure previously presented. An example of such a calculation is given by
the figure~\ref{FigChargeCentralePoly-q3p.75}, where the solid lines
show the central charge (with up and down limits of the error bars) as
a continuous function of $K$, compared to a set of independent computations
(circles) at fixed coupling strengths.

\subsubsection{Phase diagram}

\begin{figure}[ht]
	\epsfysize=9cm
	\begin{center}
	\mbox{\epsfbox{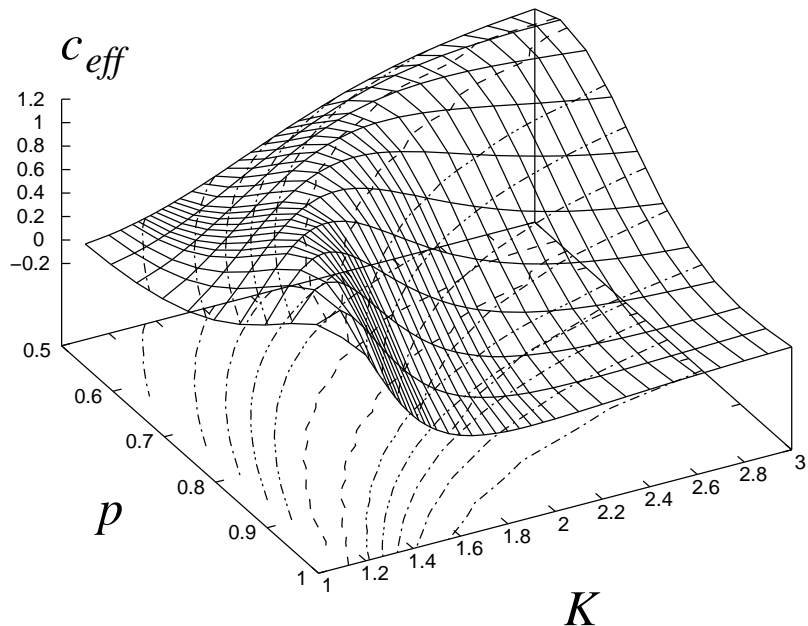}}
	\end{center}\vskip 0cm
	\caption{Dependence of the effective central charge $c_{eff}$ 
	with the exchange coupling $K$ and the bond probability $p$ for 
	the 4-state Potts model. The maximum of $c_{eff}$
	gives the location of the 
	transition line.}
	\label{Figbilan-c_eff}
\end{figure}
The phase diagram can be obtained as usually by the evolution of the maximum
of some diverging quantity, let say the susceptibility for example which
diverges at criticality in the thermodynamic limit. In the strip geometry
considered here, we preferred another technique based on the behaviour of the 
effective central charge. In usual critical systems, the 
Zamolodchikov's $c-$theorem states that there exists a $c-$function
decreasing along RG flows and giving the central charge at the fixed 
point~\cite{zamolo}.
In the case of random systems ($n\to 0$ in the replica approach), the 
central charge
increases and can be expected to reach a maximum value at an optimal
disorder amplitude. 
This property, linked to non-unitarity in the presence 
of disorder, is indeed observed in 
simulations~\cite{JacobsenCardy98,ChatelainBerche98b,ChatelainBerche99,ChatelainBerche00,DotsenkoJacobsenLewisPicco99,JacobsenPicco00}.
This property is illustrated in figure~\ref{Figbilan-c_eff} where we can
follow the maximum of the effective central charge in the plane $(p-K)$.

\subsubsection{Correlation Functions}

The spin-spin correlation functions along the strip are
calculated using an extension of the Hilbert space that allows to keep track
of the connectivity with a given spin.
For a specific disorder realisation, the spin-spin correlation function
along the strip 
\begin{equation}
	G_{\sigma}(\tau)=\frac{q\langle\delta_{
	\sigma_j\sigma_{j+\tau}}\rangle-1}{q-1},
	\label{eq-Gu}
\end{equation}	
 where $\langle\dots\rangle$ denotes the 
thermal average, is given by the probability that the spins along some row,  
at columns $j$
and $j+\tau$, are in the same state and is expressed in terms of a product
of the non-commuting transfer matrices:	
\begin{equation}
	\langle\delta_{\sigma_j\sigma_{j+\tau}}\rangle=\frac{
	\langle 0\!\mid{\bf g}_j
	\left(\prod_{k=j}^{j+\tau-1}
	{\bf T}'_k\right){\bf d}_{j+\tau}\mid\! 0\rangle}{
	\langle 0\!\mid
	\prod_{k=j}^{j+\tau-1}
	{\bf T}_k\mid\! 0\rangle},
\label{eq-corr}
\end{equation}
where $\mid\! 0\rangle$ is the ground state eigenvector, ${\bf T}_k'$ is the 
transfer matrix in the extended Hilbert space. The operator 
${\bf g}_j$ identifies the cluster containing $\sigma_j$, while
${\bf d}_{j+\tau}$ gives the appropriate weight depending on whether or not
$\sigma_{j+\tau}$ is in the same state as $\sigma_j$.
They were computed on strips of widths $L=2$ to 8 and then averaged over
$100\ 000$ disorder realisations.

\begin{figure}
	\epsfysize=10cm
	\begin{center}
	\mbox{\epsfbox{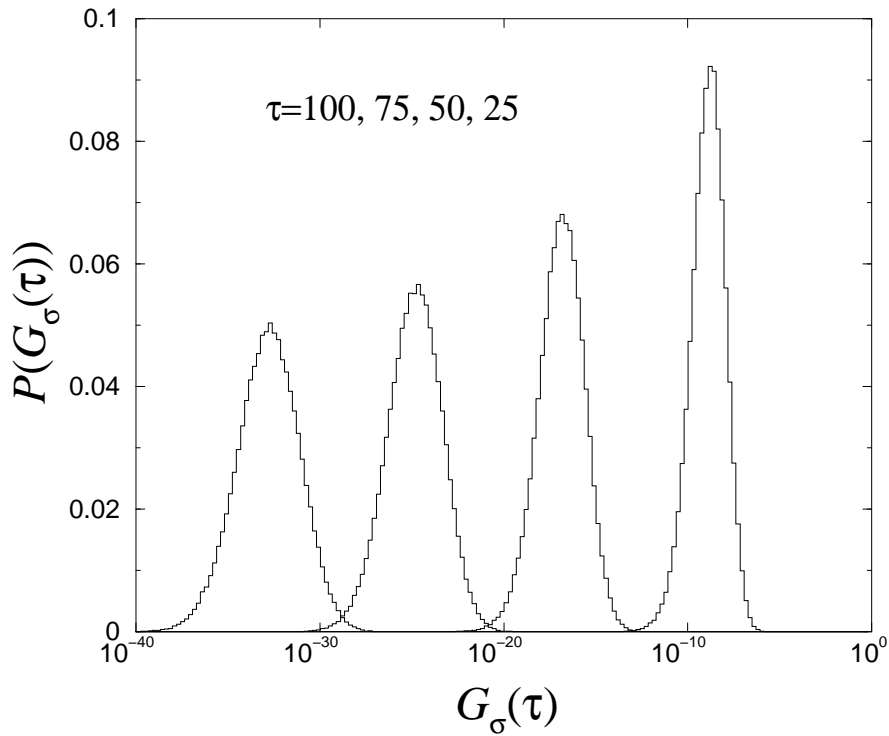}}
	\end{center}\vskip 0cm
	\caption{Histograms of probability distributions of the spin-spin 
		correlation
		function at given distances $\tau=100$, 75, 50 and 25 (from
		left to right) for the 4-state Potts model
		with an exchange coupling $K=1.16215$ and at dilution
		$p=0.75$. Note that the scale is logarithmic on the $x$-axis.}
	\label{FigHisto-G}
\end{figure}

\begin{figure}
	\epsfxsize=9.5cm
	\hfill\mbox{\epsfbox{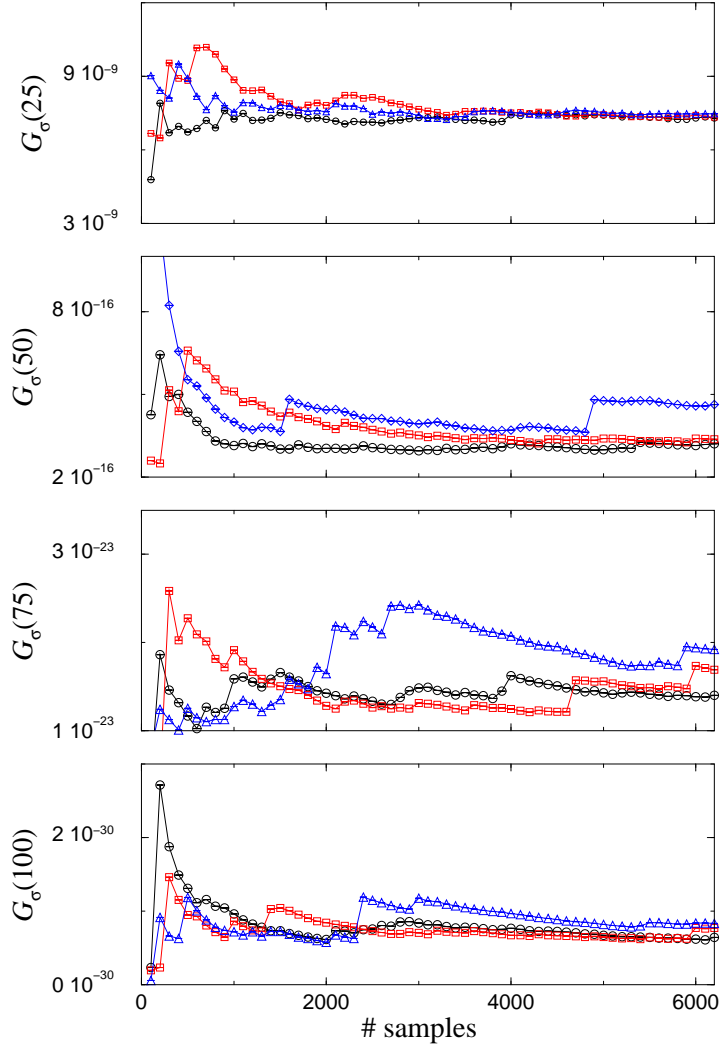}}\hfill\ 
	\caption{Convergence of the spin-spin 
		correlation functions $G_\sigma(\tau)$ at a given 
		distance $\tau=100$, 75, 50 and 25 (from bottom to top)
		with respect to 
		the number of disorder realisations (up to 6200 samples).
		}
	\label{FigConvergence-G}
\end{figure}

The numerical calculation of the average over randomness of these spin-spin
correlation functions is made difficult by the fact that they are not
self-averaging. Indeed,
Figure~\ref{FigHisto-G} shows that the probability distribution is close to a
log-normal distribution (the correlation function is given by a product
of matrices whose elements are reminiscent of randomness). 
There are thus some rare
events $G_\sigma(\tau)\sim {\cal O}(1)$ with a large relative contribution to 
the average. This is especially true at large distances. 
An accurate estimation of the average requires that the sampling
includes such events.
As can be seen in figure~\ref{FigConvergence-G},
the running average of the spin-spin correlation
function presents jumps due to these rare events. The error bars are
underestimated as the standard deviation of the correlation 
functions (which cannot correspond to
the true definition of the error on data distributed according to a
clearly non Gaussian distribution). 

\begin{figure}
	\epsfxsize=9.5cm
	\hfill\mbox{\epsfbox{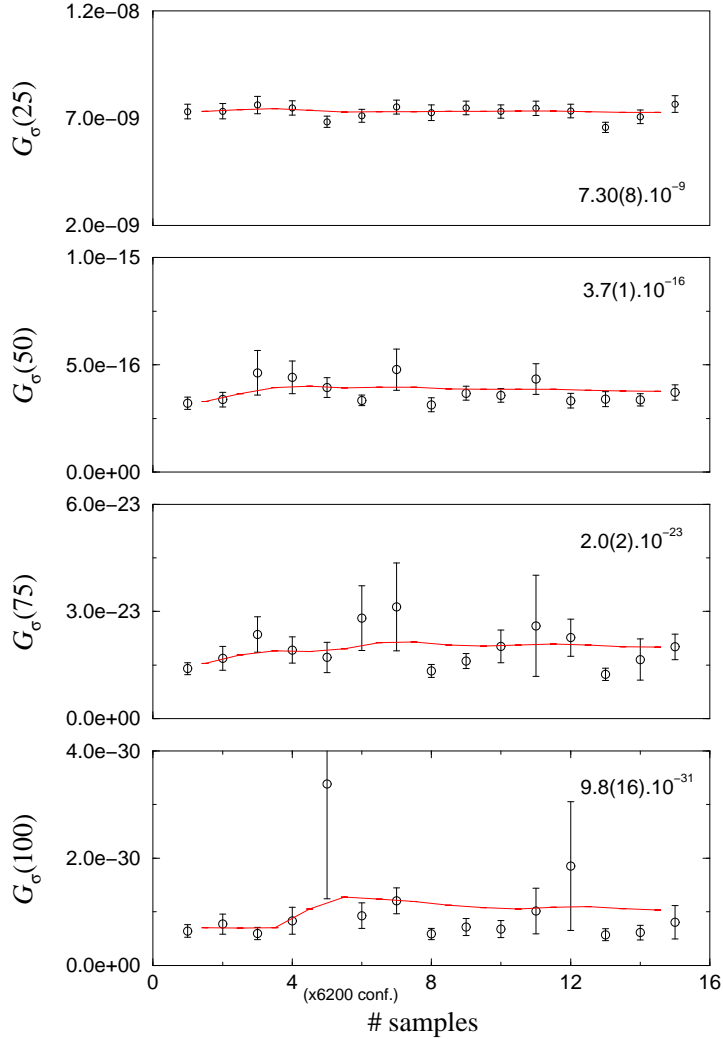}}\hfill\ 
	\caption{Convergence of the average of the spin-spin correlation
		functions $G_\sigma(\tau)$ at a given distance $\tau$ with 
		respect to the number
		of bins of $6200$ disorder realisations.}
	\label{FigConvergence-GAvg}
\end{figure}

\begin{figure}
	\epsfysize=10cm
	\begin{center}
	\mbox{\epsfbox{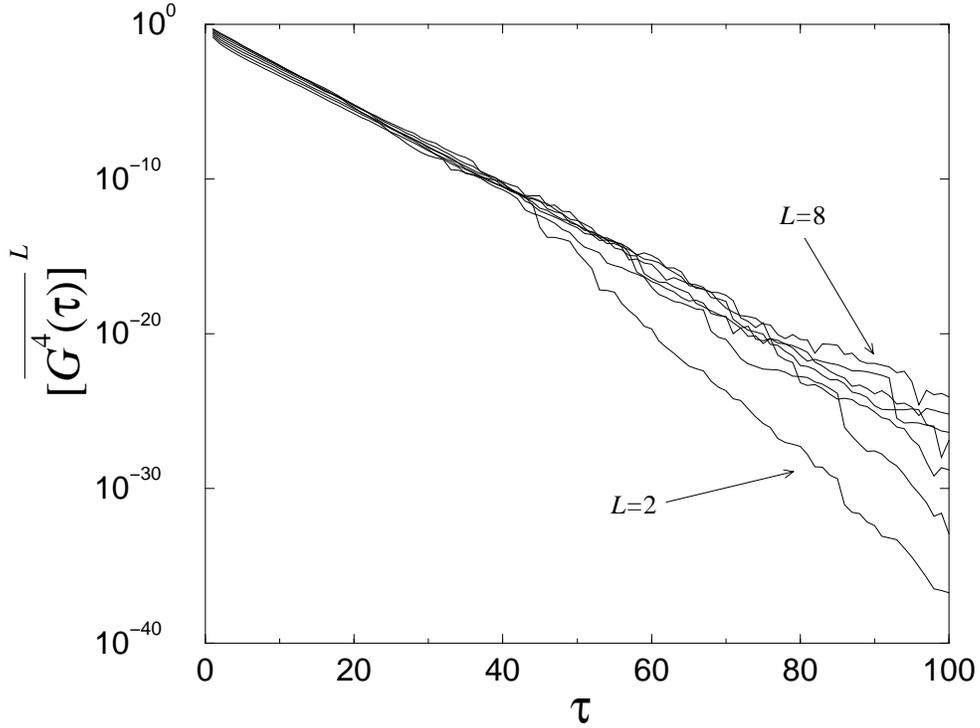}}
	\end{center}\vskip 0cm
	\caption{Jumps in the $4^{\rm th}$-moment of the spin-spin correlation
		 functions with respect to the distance at $q=4$, $p=0.70$ and 
		 $K=0.81$. The different curves correspond to lattice sizes
		 ranging from $2$ to $8$ and the average is performed over
		100~000 samples. The power $L$ ensures the same asymptotic 
		slope for different strip widths.} 
	\label{FigAnnulation-G}
\end{figure}

In the following, results are obtained using an average over 100000 disorder
realisations. As shown in figure~\ref{FigConvergence-GAvg}, it seems that
the rare events have been sufficiently well sampled with this 
number of disorder realisations, since the running average remains flat and
no systematic deviation is seen. 
For some disorder realisations, 
there can be one or several rows with no bond
at all. 
Such cuts disconnect the strip in
two subsystems and the correlation function is thus vanishing for
larger distances for the corresponding realisations, inducing a 
discontinuity of $G_\sigma(\tau)$. The average 
being performed over a finite number of
disorder realisations, these cuts lead to jumps in the average correlation
functions (figure~\ref{FigAnnulation-G}).
These jumps are more pronounced for
moments $\overline{G^n(\tau)}$ of increasing order $n$ because the contribution
of the rare events is enhanced and the average is determined by a few
configurations that can include such jumps. The averaged moments are also
less fluctuating (with the distance) for larger strip widths. 

\subsection{Methodology used in the computations}
For each value of the number of states, $q$, a few runs are performed with
$100\times 10^6$ iterations in order to have at each strip width an
evaluation of the free energy density $\overline{f}(K,p)$. Then, after
extrapolation at $L\to\infty$, the effective central charge $c_{eff}(K,p)$
is evaluated in the temperature-dilution plane.
This leads to a first approximate determination of the phase diagram $K_c(p)$.
For each dilution $p$, new runs are performed with the semi-analytical
algorithm (see section~\ref{subsubsec:improvement}) at $K_{max}$, 
the coupling strength corresponding to the maximum of
$c_{eff}$. A total of $500\times 10^6$ iterations of the 
transfer matrix are thus used.
When the curve $c_{eff}(p)$ is very flat, 
the computation is done for two distinct
values of $p$.
The spin-spin correlation functions are then computed at the maximum of
$c_{eff}(p)$ with $500\ 000$ configurations. 

\section{Phase diagram of the diluted $q$-state Potts model}
\label{sec:PhaseDiag}
\begin{figure}
	\epsfysize=10.5cm
	\begin{center}
	\mbox{\epsfbox{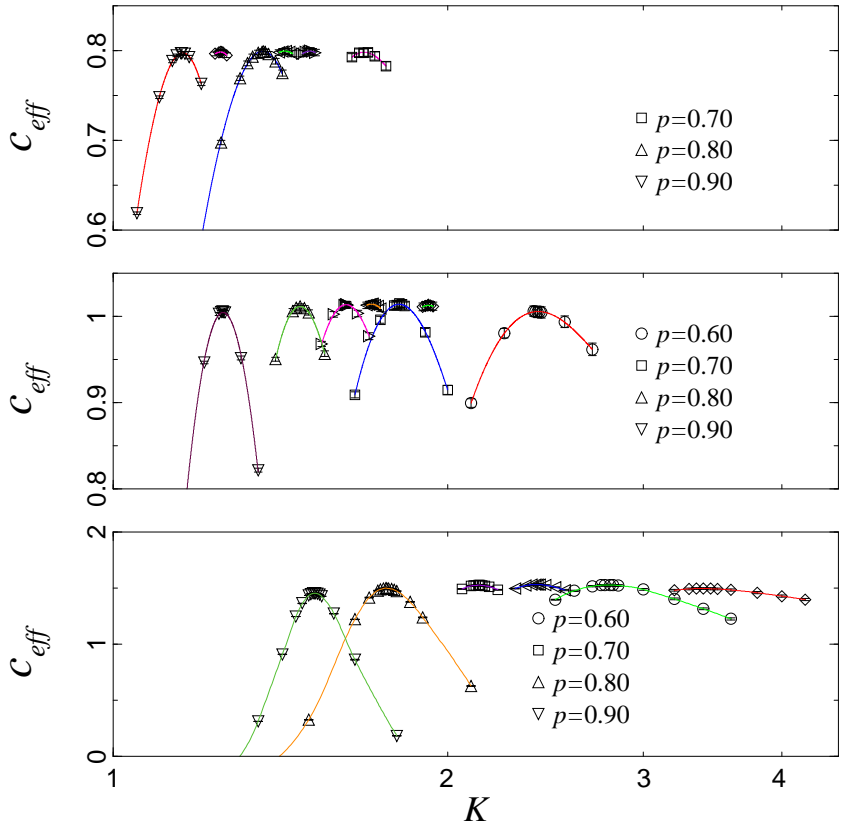}}
	\end{center}\vskip 0cm
	\caption{Estimation of the central charge with respect to the exchange
		coupling for several dilutions (for $q=3$ (top), $q=4$
	(middle) and $q=8$ (bottom)).} 
	\label{FigChargeCentrale-q348}
\end{figure}

According to the previous section, the phase diagrams of diluted Potts models
with $q=3$, $4$ and $8$ states per spin are first determined by the location
of the maxima of the central charge in strip geometries in the
$(p,K)$ plane. The effective central charge 
is shown in figure~\ref{FigChargeCentrale-q348}
for several dilutions, as a function of $K$.

\begin{figure}
	\epsfysize=10cm
	\begin{center}
	\mbox{\epsfbox{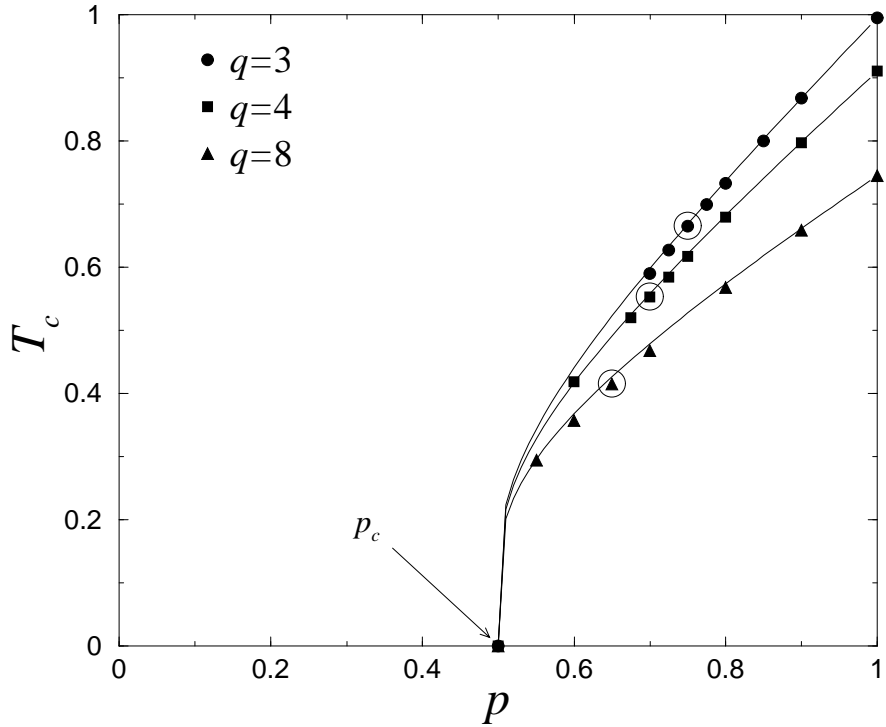}}
	\end{center}\vskip 0cm
	\caption{Phase diagram of the $q$-state diluted Potts model obtained
		numerically (symbols)  and compared  with the single-bond
	effective medium approximation (solid lines). The open circles on each
	critical line give the loci of the optimal dilutions
	which will be discussed later.} 
	\label{FigPhaseDiag348}
\end{figure}

The phase diagrams of quenched bond disordered Ising and Potts models were
studied more than twenty years ago using effective-medium 
approximation~\cite{Turban80a,Turban80b}.
The Hamiltonian is written
	\begin{equation}
	-\beta{\cal H}=\sum_{(i,j)} K_m\delta_{\sigma_i,\sigma_j}
	+\sum_{(i,j)}x_{ij}\delta_{\sigma_i,\sigma_j}
	\end{equation}
where $x_{ij}=K_{ij}-K_m$ is the deviation from the effective medium
homogeneous coupling strength. Using the identity
${\rm e}^{x_{ij}\delta_{\sigma_i,\sigma_j}}=
1+\delta_{\sigma_i,\sigma_j}({\rm e}^{x_{ij}}-1)$,
one gets a formal exact expression for the thermal average of any quantity
$Q$ as
\begin{equation}
	\langle Q\rangle=\frac{\langle Q\prod_{(i,j)}[1+
	\delta_{\sigma_i\sigma_j}({\rm e}^{x_{ij}}-1)]\rangle_m } 
	{\langle \prod_{(i,j)}[1+
	\delta_{\sigma_i\sigma_j}({\rm e}^{x_{ij}}-1)]\rangle_m} 
\end{equation}
where $\langle\dots\rangle_m$ stands for the average with Boltzmann
factors ${\rm e}^{-\beta {\cal H}_m}$ with 
\begin{equation}
	-\beta{\cal H}_m=\sum_{(i,j)} K_m\delta_{\sigma_i,\sigma_j}.
\end{equation}
A single bond approximation ($x_{ij}=0$ everywhere except on one particular 
bond) then leads to the equation of the critical line
\begin{equation}
	K_c(p)=\ln\frac{(1-p_c) {\rm e}^{K_c}-(1-p)}{p-p_c}.
\end{equation}
This expression should be exact in the vicinity of both the pure system and
the percolation threshold. Inserting the critical coupling for the pure 
system $K_c=\ln(1+\sqrt{q})\equiv K_c(p=1)$ and the percolation threshold 
$p_c=1/2$ of bond percolation on
the square lattice indeed leads to an excellent agreement with the numerical
data (see figure~\ref{FigPhaseDiag348}). We also note that the approximation 
has been improved by a cluster
extension of the effective interaction~\cite{GuilminTurban80}.

\section{Critical behaviour}\label{sec:TM}
\subsection{Critical behaviour of spin-spin correlation functions}
If the assumption of the existence of a unique stable  random fixed point 
holds, one
expects that the critical behaviour is asymptotically  the same as the 
system is moved 
along the transition
line $p_c<p<1$. However, in finite systems, one generically has to deal with
strong crossover effects due to the competition between the disordered
fixed point and the pure (at $p=1$) and percolation (at
$p=p_c$) fixed points, or to corrections to scaling linked 
to the appearence of irrelevant scaling variables. 
It is known that these latter
effects are generally important in random systems and that the
corresponding corrections to scaling can be
substantially reduced when one measures the critical exponents
in the regime of the random fixed point,  expected to be reached 
at the vicinity of the maximum of the effective central charge in the 
$p$-direction (as shown below). 
Let us consider the finite-size behaviour of an observable $Q$
measured at a deviation $t=K-K_c(p)$ from the critical point on some
system of characteristic size $L$, in the presence of dilution at bond
probability $p$. The variables $t$ and $L^{-1}$ play the role of relevant
scaling fields (with positive RG eigenvalues $y_t=1/\nu$
and $y_L=1$ respectively), while close to the fixed point, dilution is supposed
to be related to some irrelevant scaling variable with eigenvalue
$y_p=-\omega<0$. At the fixed point there is no need that the irrelevant
scaling field vanishes, so that one can write $p^*$ the corresponding
dilution and the observable $Q$ obeys the following homogeneity assumption
in the scaling region 
\begin{equation}
	Q(t,L^{-1},p)=L^{-x_Q}f(L^{1/\nu}t,L^{-\omega}(p-p^*))
\end{equation}
which corresponds to the same critical behaviour for any value of $p$
in the range $p_c<p<1$, 
described by a unique fixed point. 
An expansion of the last variable (keeping the leading term only) 
along the critical
line (i.e. varying $p$ at $K_c(p)$) gives
\begin{equation}
	Q(0,L^{-1},p)=\Gamma_QL^{-x_Q}(1+\Gamma_Q^{(2)}(p-p^*)L^{-\omega}+
	\dots),
\end{equation}
where the $\Gamma_Q$'s are non-universal critical amplitudes.
It is thus possible to fix $p=p^*$ in order to minimize the
corrections to scaling, and the corresponding value of the dilution is
empirically found to coincide with the location of the maximum of the central
charge along the critical line~\cite{ChatelainBerche99}. Close to the maximum,
the variations of the effective central charge itself are small, illustrating
that corrections to scaling have there a small influence which is consistent
with our choice in equation~(\ref{eq-fbar}).
The value of $p^*$ is not universal and should depend on the system shape,
boundary conditions, etc.

\begin{figure}
	\epsfysize=10.5cm
	\begin{center}
	\mbox{\epsfbox{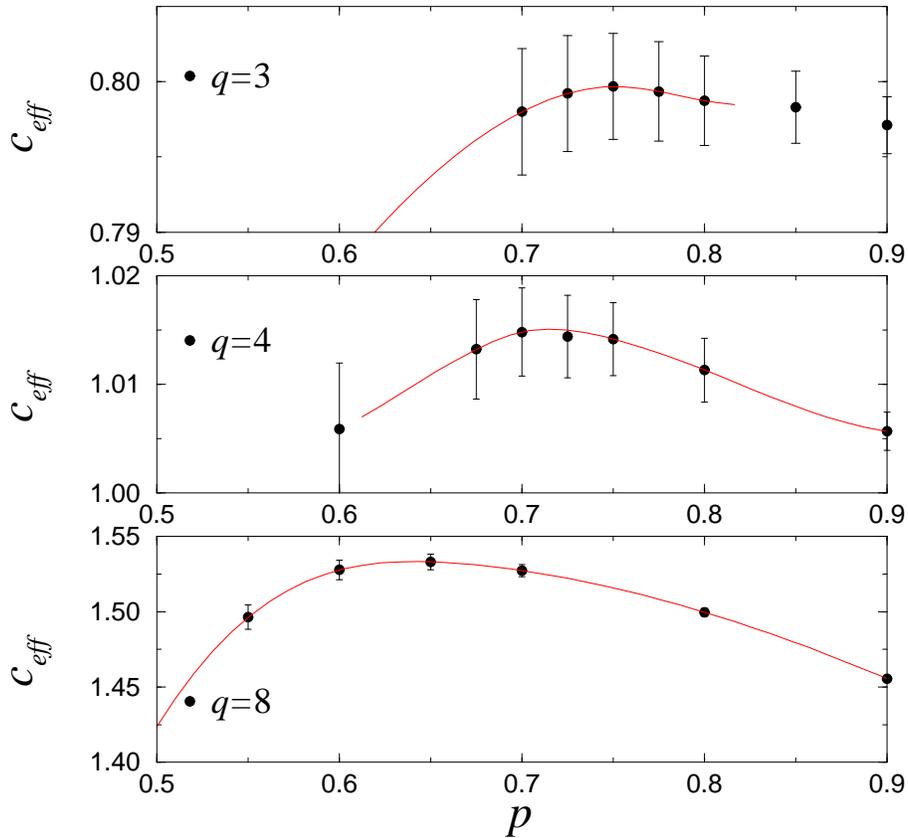}}
	\end{center}\vskip 0cm
	\caption{Estimation of the dilution $p^*$, where the central charge
	takes its maximum value. The data points correspond to
	computations performed at different temperatures, along the critical 
	line, determined by the value of the probability $p$. 
	The solid lines are simple guides for the 
	eyes.} 
	\label{MaxChargeCentrale-q348}
\end{figure}

In figure~\ref{MaxChargeCentrale-q348}, we show the 
variation of the effective central charge along the transition line. 
For example in the case $q=4$, the random fixed point corresponds roughly to 
the optimum dilution $p^*\sim 0.700$. 
The estimate of the central charge at this random
fixed point is $c^*=1.0148(40)$ which is less accurate but perfectly compatible
with the one obtained in the random-bond case with a binary 
distribution of coupling strengths ($c^*=1.0148(4)$)~\footnote{We note that a 
direct comparison between the parameters of dilution ($p$ and $K_c(p)$) with 
those of the random-bond problems is not possible. In the self-dual 
binary random-bond case for example, two couplings $K$ and $rK$ are
randomly distributed, leading to a critical line with parameter $r$, 
$K_c(r)$.}. 
This value has been
refined using the semi-analytical transfer matrix presented before and a 
5-time larger
statistic, leading to $c^*=1.0143(19)$ at $K_c^*=1.8072$. Here and in the
following, a star superscript is used to mark the values of the parameters
which lead to the maximum of $c_{eff}(K,p)$.

Once the optimal values $K_c^*$ and $p^*$ are located, we compute the correlation
functions at the corresponding point in the parameter space.
An effective magnetic critical exponent $x_\sigma(L)$ for a given strip width
$L$ is obtained by fitting the average spin-spin correlation functions with
the ansatz
\begin{equation}
        \overline{G_\sigma(\tau)}={A}\exp \left(-\frac{2\pi}{L} 
	x_\sigma(L)\tau\right)
        \label{Eq1}
\end{equation}
deduced from the logarithmic  conformal transformation which maps the
infinite plane onto an infinitely long cylinder. Here, $\tau$ is the 
coordinate along the cylinder (strip with periodic boundary conditions in 
one direction).
Conformal mappings of average profiles and correlation functions were shown to 
allow accurate determination of the critical exponents in
disordered systems, 
at least in the case of the 
random-bond Potts model~\cite{ChatelainBerche99}, and we naturally assume
here that such transformations also apply in the diluted problem. 

The data for $q=3$, 4, and 8 are summarized in table~\ref{Table1}, where our
best estimate is written in bold face. We can notice that although 
the critical exponents of the diluted and random-bond systems are quite close
as expected, the rather small error bars are probably underestimated, and 
may not reflect the true deviation when it is due to
insufficient disorder average. 

The $L$-dependence of the magnetic critical exponent
is due to corrections to scaling (due for example to the above-mentioned
crossover effects) which are not explicitly taken into account
in~(\ref{Eq1}). These effective magnetic critical exponents
$x_\sigma(L)$ are then extrapolated in the limit $L\rightarrow +\infty$ using
a polynomial of degree $2$ in $1/L$. Strip widths between $3$ and $8$ have
been used. The critical exponent associated to the decay of the average
spin-spin correlation function is for example estimated 
to be $x_\sigma\simeq 0.1419(1)$
in the case $q=4$.
This estimate is above  (but outside error bars) the estimate obtained
for the 4-state Potts model in the random-bond case ($0.1385(3)$) in the
regime of the random fixed point. We also note that a very close value is 
obtained at the critical
temperature for the dilution $p=0.725$ ($x_\sigma=0.1412(1)$).  

In the case $q=8$,
we obviously expect the first-order 
phase transition of the pure 8-state Potts model to
be turned into a second order one under the influence of
randomness~\cite{AizenmanWehr89}. We indeed  observed an exponential decay of
the spin-spin correlation functions in the longitudinal direction of the strip
with a correlation length diverging with $L$, which supports this hypothesis. 
The phase-diagram is qualitatively the same as in the case of the 4-state
Potts model. The regime
of the random fixed point corresponds in this case to 
the dilution $p^*\simeq 0.65$.
For the random-bond Potts model, the central charge in the regime of the
random fixed point was $c^*=1.5300(5)$ which is roughly what is obtained 
in the diluted case at
the maximum of the central charge $c^*=1.5320(23)$ at dilution $p^*=0.650$ 
(figure~\ref{MaxChargeCentrale-q348}). A calculation using the
semi-analytical transfer matrix yields the estimate $K_c^*=2.40420$ for the
critical exchange coupling. Unfortunately, the calculation of the spin-spin
correlation functions is made difficult by the large weight of rare events
leading to decays of correlation functions presenting jumps when the number of 
disorder realisations is too small (as 
already mentioned at the end of section~\ref{sec:algo}). Using 500,000 
disorder realisations, the decay of the average
correlation functions yields the critical 
exponent $x_\sigma=0.1514(2)$ which is
close, although again outside  error bars, to the 
estimate in the random-bond case ($0.1505(3)$).

        \begin{table}
        \caption{Central charge and magnetic critical exponent of the $q-$state
	diluted Potts model compared to the random bond Potts model at 
	the corresponding random fixed point.}
        \label{Table1}
	\begin{indented}
	\item[]
        \begin{tabular}{llllll}
	\br
	& \centre{5}{Diluted systems}  \\
	& \crule{5}  \\
	$q$ & $p$ & $K_c$ & $c$ & $x_\sigma$ & $\#$ of samples \\
	\mr
	3 & 0.750    & 1.5004     &    0.800(4) & & $100\times 10^6$ \\
	  & 0.775    & 1.4300     &    0.799(3) & & $100\times 10^6$ \\
	  &\bf 0.750 &\bf 1.50165 &\bf 0.799(2) &\bf 0.13495(6)
					& $500\times 10^6$ \\
	  & 0.775& 1.42740 & 0.799(2) & 0.13522(6)
					& $500\times 10^6$ \\        
	& \crule{5}  \\
	4 & 0.700& 1.8100  & 1.0148(40)& & $100\times 10^6$ \\
	  & 0.725& 1.7120  & 1.0144(40)& & $100\times 10^6$ \\ 
	  &\bf 0.700&\bf 1.8072  &\bf 1.0143(19)&\bf 0.1419(1)
					& $500\times 10^6$ \\  
	  & 0.725& 1.7075  & 1.0140(18)&0.1412(1)
					& $500\times 10^6$ \\  
	& \crule{5}  \\
	8 &\bf 0.650&\bf 2.40420 &\bf 1.5320(23)&\bf 0.1514(2)  
					& $500\times 10^6$ \\  
	\br
	& \centre{5}{Random Bond systems} \\
	& \crule{4}  \\
	$q$ & \centre{2}{distribution}  & $c$ & $x_\sigma$ \\
	\mr
	3   & \centre{2}{binary$^{\rm a}$\hfill\ }  & 0.7998(4) & 0.1347(11)\\
            & \centre{2}{ternary$^{\rm a}$\hfill\ } &           & 0.1344(8) \\
            & \centre{2}{quaternary$^{\rm a}$\hfill\ } &        & 0.1343(6) \\
	    & \centre{2}{continuous$^{\rm a}$\hfill\ } &        & 0.1344(13) \\
	& \crule{5}  \\
	4   & \centre{2}{binary$^{\rm b}$\hfill\ }  & 1.0148(4) & 0.1385(3)\\
	& \crule{5}  \\
	8   & \centre{2}{binary$^{\rm b}$\hfill\ }  & 1.5300(5) & 0.1505(3)\\
	\br
        \end{tabular}
	\item[] $^{\rm a}$ From \paper{C. Chatelain and B. Berche}{2000}
	{{\it Nucl. Phys. B}}{572}{626} 
	\item[] $^{\rm b}$ From \paper{C. Chatelain and B. Berche}{1999}
	{\PRE}{60}{3853}
        \end{indented}
	\end{table}

In the case $q=3$, 
the procedure is again identical to that of the 4 and 8-state Potts model 
presented before. The peak of the central charge with respect to the exchange
coupling are narrower than for larger values of $q$
(figure~\ref{FigChargeCentrale-q348}). This leads to a better defined phase
diagram (figure~\ref{FigPhaseDiag348}) 
than for $q=4$ and $q=8$. Nevertheless,
the error bars on the central charge are of the order of magnitude of its
variation with the dilution $p$. It is thus difficult to define precisely
its maximum, approximately located at $p^*=0.75$.
The effective critical exponent, as given by the 
decay of the spin-spin correlation, depends much more on the precision on the
critical exchange coupling $K_c^*$ than for larger value of $q$. The
semi-analytical transfer matrix 
at the dilution $p=0.75$ with an 
average of the free energy over $500$ products 
of $10^6$ iterations of the transfer matrix gives the refined estimate
$K_c^*\simeq 1.50165$. The maximum of the central charge is $0.799(2)$
which is compatible, within error bars, with the value $0.7998(4)$ obtained in 
the random bond case. The estimate of the magnetic critical exponent at this
point $0.13495(6)$ is compatible with the value obtained in the random-bond
case $0.1347(11)$ and with the estimate $0.13465$
obtained by perturbative developments in the neighborhood of $q=2$. 

\subsection{Multifractal behaviour of the spin-spin correlation function}
In this section, we report a study of the multifractal properties of the
spin-spin correlation functions of the diluted Potts model. A
similar analysis was performed in the 
random-bond case in Ref.~\cite{ChatelainBerche00}. The aim is to provide
in the diluted case also a test of replica symmetry breaking using the 
multifractal behaviour of the spin-spin correlation functions, then to
compare the multifractal spectrum for different values of $q$. 

The critical 
exponent
$x_{\sigma^2}$ associated to the algebraic decay of the second moment 
$\overline{G^{2}_\sigma(\tau)}^{1/2}$ of the
spin-spin correlation function was calculated using a perturbation
expansion around the conformal field theory at $q=2$ with the assumption
that replica symmetry holds or is spontaneously broken.
At $q=3$, the expansion based on the replica symmetric scenario 
was shown to be in a very good agreement with the numerical data for
the random-bond problem. In the diluted case, we confirm this
agreement, as can be seen in table~\ref{Table2}.

        \begin{table}
        \caption{Magnetic critical exponent of the second moment of the
	spin-spin correlation function of the $3-$state
	diluted Potts model compared to the random bond Potts model at 
	the corresponding random fixed point. The perturbative results
	of Dotsenko et al. are recalled.}
        \label{Table2}
	\begin{indented}
	\item[]
        \begin{tabular}{lll}\br
	\centre{3}{Transfer matrix}\\
	\crule{3}\\
	Randomness & distribution & $x_{\sigma^2}$ \\
	\mr
	Dilution & & 0.1184(1) \\
	Random bond & binary & 0.1177(12)$^{\rm a}$ \\
	& ternary & 0.1182(12)$^{\rm a}$ \\
	& continuous & 0.1173(14)$^{\rm a}$ \\
	\br
	\centre{3}{Perturbation}\\
	\crule{3}\\
	\centre{2}{Replica symmetry\hfill\ } & 0.11761$^{\rm b}$ \\
	\centre{2}{Replica symmetry breaking\hfill\ } & 0.12011$^{\rm b}$ \\
	\br
        \end{tabular}
	\item[] $^{\rm a}$ From \paper{C. Chatelain and B. Berche}{2000}
	{{\it Nucl. Phys. B}}{572}{626} 
	\item[] $^{\rm b}$ From \paper{Vik. Dotsenko, Vl. Dotsenko and M.
	Picco}{1998}{{\it Nucl. Phys.}}{B250}{633} 
	\end{indented}
        \end{table}

\begin{figure}
	\epsfysize=10cm
	\begin{center}
	\mbox{\epsfbox{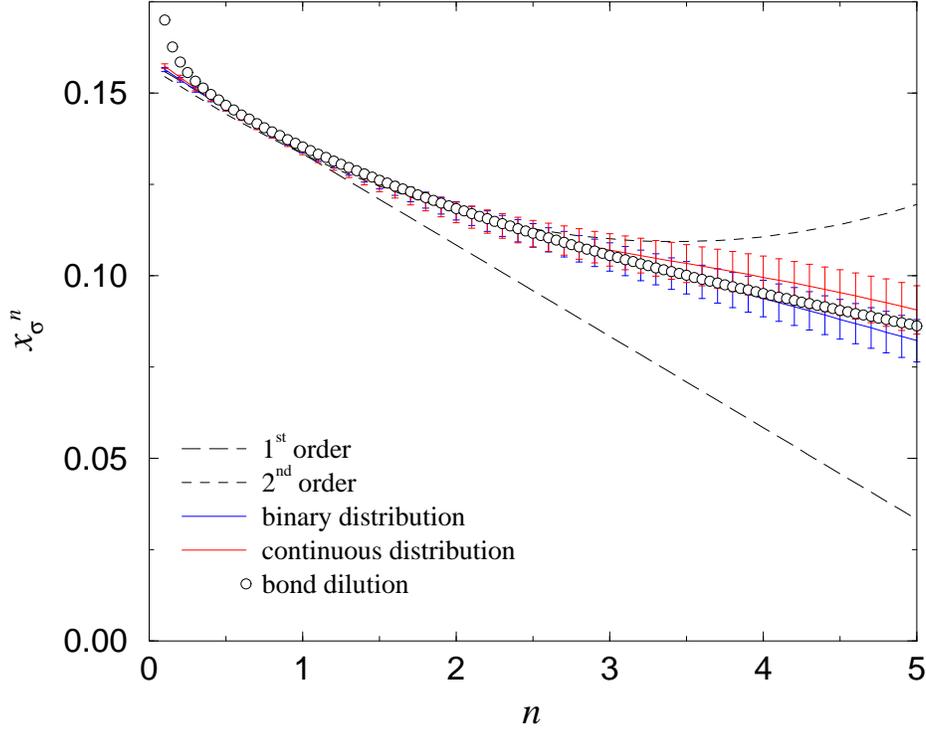}}
	\end{center}\vskip 0cm
	\caption{Exponents of the moments of the correlation function, 
	$x_{\sigma^n}$ plotted against the moment order for $q=3$.
	A convincing agreement follows from  
	comparison with 1st and 2nd order expansion calculations  
	and also with numerical results obtained for random-bond models.} 
	\label{Multifractal-q3}
\end{figure}

More generally, the critical behaviour of the moments of the correlation 
function is characterized by a set of exponents $x_{\sigma^n}$ which
depend on the moment order $n$ in the case of
multifractality. Numerically, these exponents are obtained by a simple
generalization of equation~(\ref{Eq1}) to: 
\begin{equation}
        \overline{G_\sigma^n(\tau)}^{1/n}={A_n}\exp \left(-\frac{2\pi}{L} 
	x_{\sigma^n}(L)\tau\right)
        \label{Eqn}
\end{equation}
with the extrapolation to $L\to\infty$.
The exponents associated to the first integer order moments are given in 
table~\ref{Table3} and compared to the corresponding random-bond values
and to the perturbative results in the replica symmetric 
scenario~\cite{Lewis98}. The plot is shown in figure~\ref{Multifractal-q3}.

        \begin{table}
        \caption{Magnetic critical exponent of the moments of the
	spin-spin correlation function of the $3-$state
	diluted Potts model compared to the random bond Potts model at 
	the corresponding random fixed point. The perturbative results
	of Lewis are recalled (the expansion is only valid for $q$ close
	to the Ising value $q=2$ and at small moment order $n$ close to 1).}
        \label{Table3}
	\begin{indented}
	\item[]
        \begin{tabular}{lllll}\br
	&\centre{4}{$x_{\sigma^n}$} \\
	&\crule{4}\\
	&&\centre{3}{Transfer matrix}\\
	&&\crule{3}\\
	$n$ & theor.$^{\rm a}$ & binary$^{\rm b}$ & continuous$^{\rm b}$ 
		& diluted \\
	\mr
	1 & 0.13465 & 0.1347(11) & 0.1344(13) & 0.13522(6) \\
	2 & 0.11761 & 0.1177(25) & 0.1173(28) & 0.11841(11) \\
	3 & 0.11006 & 0.1051(39) & 0.1070(45) & 0.10552(17) \\
	4 &    -    & 0.0938(50) & 0.0996(58) & 0.09511(21) \\
	5 &    -    & 0.0822(58) & 0.0906(66) & 0.08630(25) \\
   	\br
        \end{tabular}
	\item[] $^{\rm a}$ From M.A. Lewis, {\it Europhys. Lett.} 
	{\bf 43}, {189} 
	(1998). 
	\item[] $^{\rm b}$ From \paper{C. Chatelain and B. Berche}{2000}
	{{\it Nucl. Phys. B}}{572}{626} 
	\end{indented}
        \end{table}

An interesting quantity is given by the Legendre transform $H(\alpha)$ of
the set of exponents $X_{\sigma^n}=nx_{\sigma^n}$, 
\begin{equation}
	H(\alpha)=X_{\sigma^n}-n\alpha,\ \alpha={\partial 
	X_{\sigma^n}\over\partial n},\ n=-\frac{\partial H}{\partial\alpha}.
\end{equation}
This multifractal function is naturally introduced through the 
probability distribution of the quantity of interest.
The exponential decay of the moments of the
correlation function along the strip
\begin{equation}
        \overline{[G_\sigma(\tau)]^n}
	\equiv\int_0^1 [G_\sigma(\tau)]^nP_\tau(G){\rm d}G	
	\sim\exp \left(-\frac{2\pi}{L} 
	X_{\sigma^n}(L)\tau\right),
        \label{Eqnn}
\end{equation}
may be seen as a function $L_\tau(n)$,
which is, by construction, the Laplace transform of the probability
distribution $P_\tau(y)$ at fixed $\tau$, with the positive variable 
$y=-\ln  G_\sigma(\tau)$:
\begin{equation}
	\overline{[G_\sigma(\tau)]^n}\equiv
	L_\tau(n)=\int_0^\infty{\rm e}^{-ny}P_\tau(y){\rm d}y.
\end{equation}
Inverting the Laplace transform,
\begin{equation}
	P_\tau(y)=\frac{1}{2{\rm i}\pi}\int_{\delta-{\rm i}\infty}^{\delta
	+{\rm i}\infty}{\rm e}^{-\frac{2\pi\tau}{L}
	[X_{\sigma^n}-ny/(2\pi\tau/L)]} {\rm d}n	
\end{equation}
and performing a saddle-point
approximation of the Bromwich integral~\cite{FourcadeTremblay87},
a similar exponential expression follows for the
probability distribution 
\begin{equation}
        P_\tau(y)\sim
	\exp \left(-\frac{2\pi}{L} 
	H(\alpha)\tau\right)
        \label{Eqpdey}
\end{equation}
where the variable $\alpha=-\frac{L}{2\pi \tau}\ln G_\sigma(\tau)$.
We implicitly make use of the assumption that the amplitude in 
equation~(\ref{Eqnn})
only smoothly depends on $n$\footnote{$a_n$ being the 
amplitude of the
$n^{th}$ order moment, the term $-\ln a_n/(2\pi\tau /L)$ vanishes in the
limit of large strips $\tau /L\gg 1$ and can be forgotten in the inverse
Laplace transform.}. 
If the scaling dimensions $X_{\sigma^n}$ measure the critical decay of the
correlation functions (or more generally of the moments of the correlation
function), the spectral function 
$H(\alpha)$ 
measures the exponential decay
of the correlation function probability distribution, and it is absolutely
equivalent to work in terms of scaling dimensions or in terms
of spectral function.
The latter quantity is also the natural universal scaling function which
allows a rescaling of the probability distributions of the
correlation functions at different distances along the strip or with
different strip widths.
It is obtained using the identity 
$P_\tau(\alpha)=(2\pi\tau/L)P_\tau(\ln G_\sigma(\tau))$ and
$	H(\alpha)=-\frac{L}{2\pi\tau}\ln P_\tau(\alpha)$,
up to an additional term which does not depend on $\alpha$, but does depend
explicitly on $2\pi\tau/L$, and comes from the change of variable from
$y$ to $\alpha$ and on the possible
refinement of the saddle-point approximation.

The maximum of the probability distribution $P_\tau(\alpha)$ corresponds
to the average $\overline{\ln G_\sigma(\tau)}$ and
defines the typical (or most probable value) of the correlation function. 
It is obtained
at the minimum of the spectral function $H(\alpha)$ at position $\alpha_0$. 
Following the 
definition of $\alpha$,  
$\alpha_0=\bar y/(2\pi\tau/L)=-\overline{\ln G_\sigma(\tau)}/(2\pi\tau/L)$
also corresponds to the scaling dimension of the typical correlation
function.
\begin{equation}
	G_{typ}(\tau)\equiv\exp\overline{\ln G_\sigma (\tau)}=
	\exp\left(-\frac{2\pi\tau}{L}\alpha_0\right).
\end{equation}
The geometrical interpretation of the function 
$H(\alpha)$ is the following: the curve $H(\alpha)$ has in $\alpha$ a tangent 
$n(x_{\sigma^n}-\alpha)$ of slope $-n=\partial H/\partial\alpha$  
which intercepts 
the horizontal axis at $\alpha=x_{\sigma^n}$ and the vertical axis 
at $X_{\sigma^n}=nx_{\sigma^n}$. At the minimum $H(\alpha_0)$, 
$n=0$, implying at the same time that  the value of $H(\alpha_0)$ 
also vanishes.
For a log-normal distribution, the spectral function is simply parabolic,
and the deviation from the parabola measures the distance from the log-normal
probability distribution.

\begin{figure}
	\epsfysize=10.5cm
	\begin{center}
	\mbox{\epsfbox{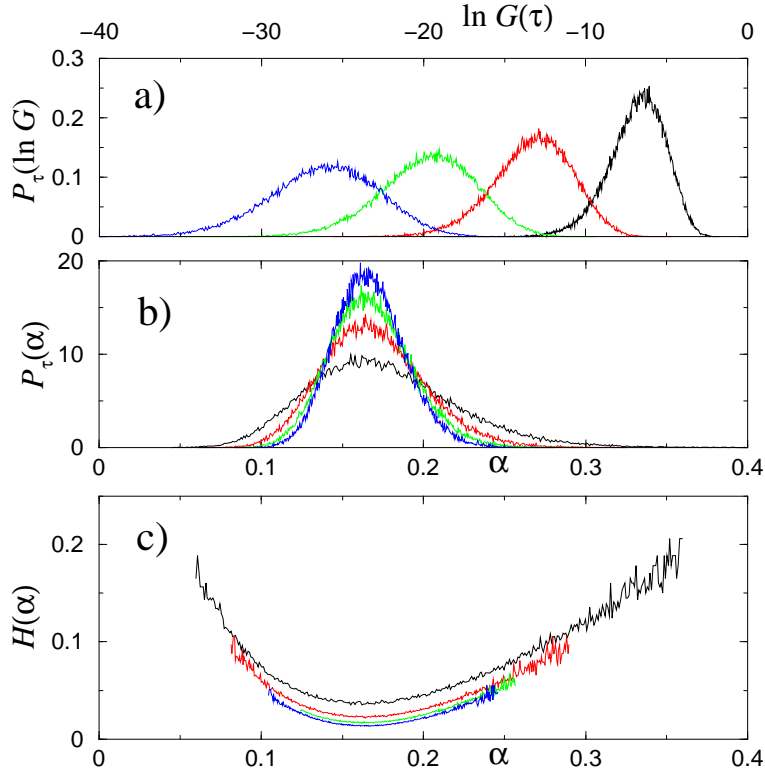}}
	\end{center}\vskip 0cm
	\caption{Probability distribution of the spin-spin correlation
	function for the 3-state diluted Potts model at criticality at
	the optimal dilution $p^*$. 
	a) Probability distribution of the
	$\ln G_\sigma(\tau)$ for different distances along the strip, 
	$\tau=50$, 100, 150 and 200 from right to left. 
	b) Horizontal rescaling using the variable 
	$\alpha=-\ln G_\sigma(\tau)/(2\pi
	\tau/L)$, with the same values of $\tau$ from top to bottom. 
	c) Rough approximation of the universal spectral 
	function which should no longer 
	depend on the distance along the strip. It is given here at 
	the saddle-point approximation,
	$H(\alpha)=-\frac{L}{2\pi\tau}\ln \left[\frac{L}{2\pi\tau}
	P_\tau(\alpha)\right]$.} 
	\label{DistProbG_q3Fig}
\end{figure}
\begin{figure}
	\epsfysize=10.5cm
	\begin{center}
	\mbox{\epsfbox{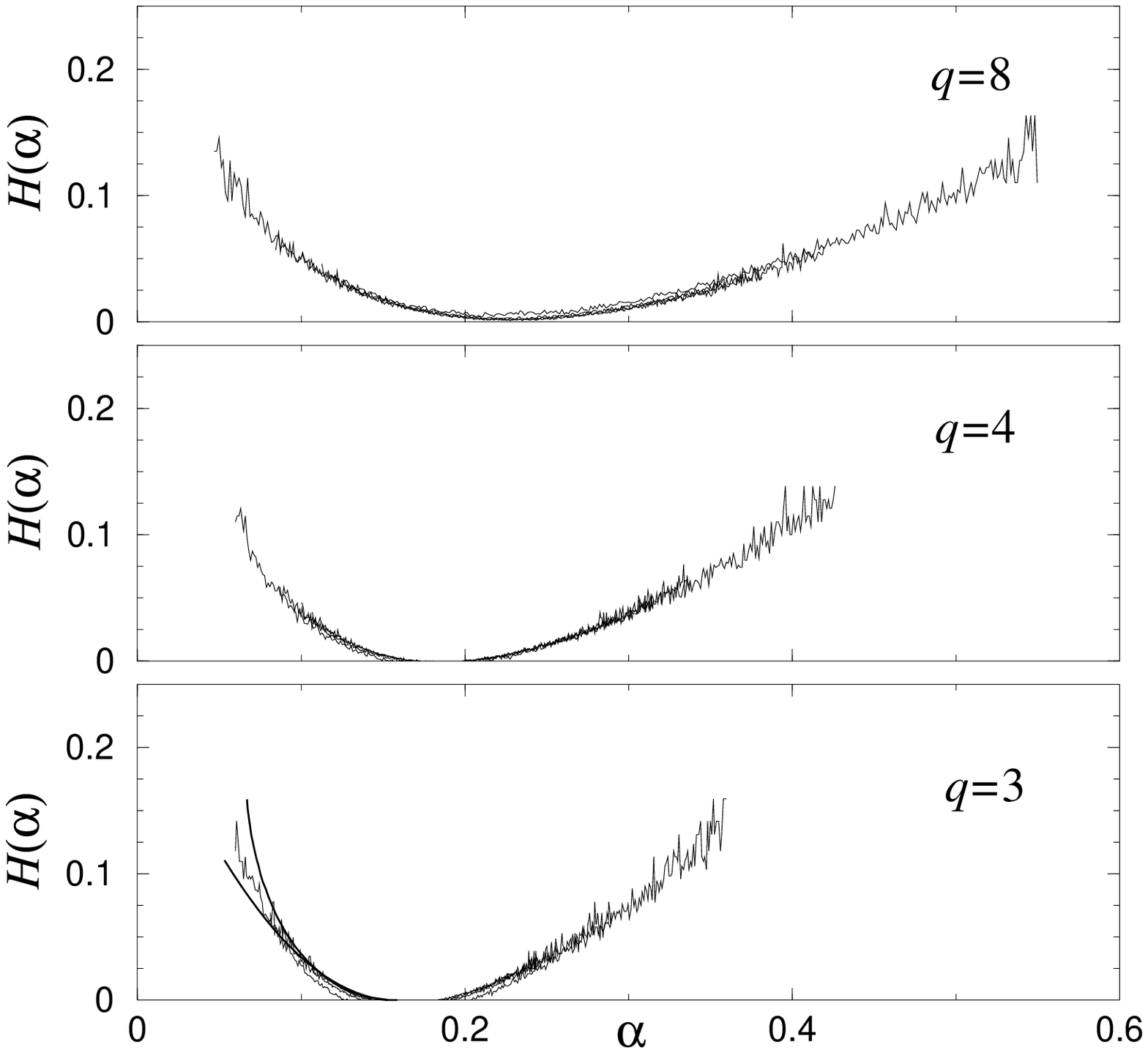}}
	\end{center}\vskip 0cm
	\caption{Spectral function $H(\alpha)$ for $q=3$, 4, and 8 
	state per spin. The data
	corresponding to distances $\tau=50$, 100, 150 and 200 collapse
	onto single universal spectral functions.
	The multifractal spectrum becomes wider as $q$ is increased.
	At $q=3$, the solid lines are the results deduced from 1st and 2nd 
	order expansions of $X_{\sigma^n}$ given in 
	Refs.~\protect{\cite{Ludwig87}} and \protect{\cite{Lewis98}}.} 
	\label{DistProbG_q348}
\end{figure}

In the case of the diluted $3-$state Potts model, the probability distribution
of the spin-spin correlation function  
at distances $\tau=50$, 100, 150, and 200
is presented in figure~\ref{DistProbG_q3Fig}
(in fact $P_\tau(\ln G)$ is shown). As one can see, the distribution
is very broad, and the broadening is more pronounced at large distances.
We mention that the events corresponding to vanishing $G_\sigma(\tau)$ have
been discarded (at $\tau=50$, it corresponds to $3\%$ of the events for $q=3$
($22\%$ for $q=8$) and this proportion increases at $\tau=200$ to almost $5\%$
for $q=3$ (and $52\%$ for $q=8$)).
A simple change to the natural variable $\alpha$ already produces a
rescaling in the horizontal direction, and the spectral function deduced
from the saddle-point approximation,
	$H(\alpha)=-\frac{L}{2\pi\tau}\ln \left[\frac{L}{2\pi\tau}
	P_\tau(\alpha)\right]$,
is shown 
in the same figure. At this approximation,
the maxima  $P_\tau(\alpha_0)$ should be given by $\frac{2\pi\tau}{L}$ (which
take values close 157, 118, 78 and 40 for the values of $\tau$ chosen here).
Since it is clearly not true (see figure~\ref{DistProbG_q3Fig} b) the minima
of $H(\alpha)$ do not vanish and the rescaling is not perfect. We thus have
to keep quadratic fluctuations about the 
saddle $n^*$~\cite{AharonyBlumenfeld93}
leading to the following estimate of the Bromwich integral
\begin{equation}
	P_\tau(\alpha)=\left[\frac{1}{2\pi}\left(\frac{\partial^2X_{\sigma^n}}
	{\partial n^2}\right)_{n^*}\right]^{-1/2}\left(\frac{2\pi\tau}
	{L}\right)^{1/2}\exp\left(-\frac{2\pi\tau}{L}H(\alpha)\right)
\end{equation}
The rescaling is indeed clearly improved with
\begin{equation}
        H(\alpha)=-\frac{L}{2\pi\tau}\ln \left[\left(\frac{L}{2\pi\tau}
	\right)^{1/2} P_\tau(\alpha)\right]
	\label{EqHdealpha}
\end{equation}
provided that the prefactor $\left[\frac{1}{2\pi}\left(\frac{\partial^2
X_{\sigma^n}}{\partial n^2}\right)_{n^*}\right]^{-1/2}$ is of order
unity. With the numbers taken from the maxima of
figure~\ref{DistProbG_q3Fig} b), we deduce that at $q=3$ this prefactor
takes a value close to $1.5$\footnote{For example for $q=3$, the prefactors
$\left[\frac{1}{2\pi}\left(\frac{\partial^2
X_{\sigma^n}}{\partial n^2}\right)_{n^*}\right]^{-1/2}$
take the values 1.48, 1.50, 1.47 and 1.44 for $\tau=50$, 100, 150, and 200,
respectively, leading to additive corrections at most
$\frac{L}{2\pi\tau}\ln\left[\frac{1}{2\pi}\left(\frac{\partial^2
X_{\sigma^n}}{\partial n^2}\right)_{n^*}\right]^{-1/2}\simeq -0.010$, 
which explains the negative shift
observed in figure~\ref{DistProbG_q348} for $q=3$, while it is even smaller
for $q=4$ (-0.003) and really negligible for $q=8$.}, 
making the correction term to 
equation~(\ref{EqHdealpha}), 
$\frac{L}{2\pi\tau}\ln\left[\frac{1}{2\pi}\left(\frac{\partial^2
X_{\sigma^n}}{\partial n^2}\right)_{n^*}\right]^{-1/2}$, indeed negligible.
This is also the case  for $q=4$ and 8 and the
spectral functions presented in figure~\ref{DistProbG_q348}
have a vanishing minimum and produce a
very good rescaling of the correlation function probability distributions. 
The multifractal
spectrum is broader at larger values of the number of states per spin,
indicating a more important relative weight of the rare events, which implies
that numerical results would become less reliable for large $q$'s.

\section{Conclusion}\label{sec:ccl}
We have studied the magnetic critical properties of the bidimensional diluted
Potts model for $q=3$, 4, and 8. 
From transfer matrix computations, we obtained the critical 
exponents of the decay of the average spin-spin
correlation function, $x_\sigma$, which are close to those obtained
in the random-bond case. 
The maximum value of the central charge is found to be compatible 
with the estimates in the corresponding random-bond models. 
Finally, the multifractal spectrum of the diluted
3-state Potts model turns out to be compatible with that obtained both
perturbatively and numerically for the random-bond case. This is a strong
evidence of the existence of an unique fixed point for all kinds of disorder,
even in the absence of duality symmetries.
In which concerns the multifractal properties, the spectral function
$H(\alpha)$ becomes broader when the number of states per spin increases,
indicating extremely stretched probability distributions of the 
correlation functions.

\ack 
This work has been supported by the
twinning research programme between Landau Institute and l'Ecole Normale
Sup\'erieure de Paris. Partial support from RFBR through Project 
99-07-18412 is also acknowledged. 
C.C. thanks the Institut f\"ur Theoretische Physik Leipzig for hospitality 
and support through a postdoctoral
position in European Network "Discrete Random Geometries: from solid
state physics to quantum gravity".

The computations  were performed using the facilities of the CINES 
in Montpellier under project No. C20010620011.

\end{document}